\documentclass[prd, aps, nobibnotes, nofootinbib,
eqsecnum,
preprint,
showpacs,
preprintnumbers,
amsmath,
amssymb]{revtex4-1}

\usepackage{graphicx}
\usepackage{bm}
\usepackage{hyperref}
\usepackage{mathrsfs}
\usepackage{xcolor}
\usepackage{subcaption}
\usepackage{multirow}
\usepackage{hhline}

\def\Tr{\mathrm{Tr}}

\begin{document}

\title{{\bf{\Large Cosmology of Bianchi type-I metric using renormalization group approach for quantum gravity}}}

\author{ Rituparna Mandal}
\email{drimit.ritu@gmail.com, rituparna1992@bose.res.in}\,
\affiliation{Department of Theoretical Sciences, S.N. Bose National Centre for Basic Sciences, Block JD, Sector III, Salt Lake, Kolkata 700106, India}
\author{ Sunandan Gangopadhyay}
\email{sunandan.gangopadhyay@gmail.com, sunandan.gangopadhyay@bose.res.in},\,
\affiliation{Department of Theoretical Sciences, S.N. Bose National Centre for Basic Sciences, Block JD, Sector III, Salt Lake, Kolkata 700106, India}
\author{ Amitabha Lahiri}
\email{amitabha@bose.res.in }
\affiliation{Department of Theoretical Sciences, S.N. Bose National Centre for Basic Sciences, Block JD, Sector III, Salt Lake, Kolkata 700106, India}


\begin{abstract}
We study the anisotropic Bianchi type-I cosmological model at late times, taking into account quantum gravitational corrections
in the formalism of the exact renormalization group flow of the effective average action for gravity.
The cosmological evolution equations are derived by including the scale dependence of Newton's constant $G$ and cosmological constant $\Lambda$. We have considered the solutions of the flow equations for $G$ and $\Lambda$ at next to leading order in the infrared cutoff scale. 
Using these scale dependent $G$ and $\Lambda$ in Einstein equations for the Bianchi-I model, we obtain the scale factors in different directions. It is shown that the scale factors eventually evolve into FLRW universe for known matter like radiation. However, for dust and stiff matter we find that the universe need not evolve to the FLRW cosmology in general, but can also show Kasner type behaviour.
\end{abstract}
\vskip 1cm
\maketitle

\section{Introduction}
General relativity is extremely successful as the theory of low energy, long distance 
gravitational interactions. The detection of gravitational waves from colliding black 
holes and neutron stars is in agreement with general relativistic predictions at an 
extraordinary level of instrumental precision, as are observations of the dynamics 
of binary pulsars, transfer of time measurements to atomic clocks in satellites of 
the Global Positioning System (GPS), and many other observations in the solar system
as well as of astronomical objects (for a review of tests of general relativity 
and the agreement between theory and experiment, see~\cite{Will:2014kxa, Will:2018bme}).
There are many formal similarities between general relativity and non-Abelian gauge 
theories, starting from the fact that both are field theories based on local symmetries
(see e.g.~\cite{Utiyama:1956sy, Weinberg:1972kfs, Hehl:1976kj, Hehl:1976my, Blagojevic:2002du, 
	Blagojevic:2013xpa})\,
Unlike non-Abelian gauge theories however, no consistent quantization of general relativity 
is known. The gravitational coupling is the dimensionful Newton's constant $G_N \sim \frac{1}{M_P^2}\,,$
while for gauge theories the coupling constants are dimensionless. So at each loop order 
in perturbative quantum gravity, the ultraviolet divergence is worse than in gauge theories 
by two powers of loop momentum. As a result, the number of distinct counterterms required to 
renormalize quantum gravity can be expected to be infinite, whereas the number of 
counterterms for a perturbatively renormalizable quantum gauge theory is finite. 

The failure of perturbative methods to consistently quantize gravity does not
however mean that a quantum theory of gravity cannot exist. It is possible 
for example that nonperturbative approaches such as loop quantum gravity~\cite{rovelli:2004, thiemann:2004}
may lead to a consistent quantum theory, or that by 
embedding gravity in a bigger theory such as supergravity~\cite{wess:1992, Bern:2009kd} 
or string theory~\cite{polchinski:1998a, polchinski:1998b, kiritsis:2007} 
it may be possible to find a consistent quantum description of gravitation 
in four spacetime dimensions. Another possibility is that general relativity 
cannot be quantized but emerges as an effective field theory at low energies, 
thus including all possible diffeomorphism-invariant local functions of the metric
which are not ruled out by other symmetries~\cite{donoghue:1994, verlinde:2011, padmanabhan:2010}.  

Yet another approach is to assume that the quantum theory which describes gravity
in four dimensions is an ``asymptotically safe'' theory, i.e. 
the essential couplings of the theory hit a fixed point as the scale at 
which they are calculated is taken to infinity~\cite{Weinberg:1980gg}. 
What this means is the following. 
Among all the couplings of the theory there are some ``inessential 
couplings'' $Z$ for which $\dfrac{\partial{\mathscr{L}}}{\partial Z}$
is either zero or a total derivative when the field equations are satisfied. 
An example is the wave function renormalization constant, which can be eliminated 
by redefining the fields. The remaining coupling constants are the essential 
couplings, which will flow with an external parameter $k$ which has the 
dimensions of mass. The meaning of this $k$ depends on 
physics of the problem -- it can be the momentum transfer in a scattering problem,
or the inverse of some length scale specific to the problem. If the essential
couplings $g_i$ are dimensionful, we make them dimensionless by multiplying with
suitable powers of $k$. Then the $k$-dependence
of the essential couplings $g_i$ are characterized by the $\beta_i(g) = k\dfrac{d g_i}{d k}$\,,
called the $\beta$-functions. In general the $\beta$-function for any of the
$g_i$ will depend on all of the essential couplings. The $\beta$-functions
of any theory describe a trajectory on the space of coupling constants.
 If a theory has a fixed point $g^*$ in 
the space of all the $g_i$\,, the beta functions must vanish at that point, and also 
the trajectory for the theory must hit the point $g^*$\,. The trajectories which 
hit the fixed point form a hypersurface called the ``critical surface'', and asymptotically
safe theories are defined to be those for which the couplings lie on the critical surface 
of a fixed point. Many non-Abelian gauge theories have ultraviolet 
fixed points where the gauge coupling vanishes, making the theory asymptotically 
free~\cite{Gross:1973id, Politzer:1973fx}. For gravity, several calculations based on 
truncated Exact Renormalization Group Equations (ERGE) support the conjecture
that there is an interacting fixed point in the ultraviolet regime~\cite{Reuter:1996cp, 
	Lauscher:2001ya, Lauscher:2002sq, Souma:1999at, Percacci:2005wu, Niedermaier:2006wt, 
	Niedermaier:2006ns, Niedermaier:2010zz, Falls:2014tra}. Support for the conjecture also comes from 
calculations for gravity with matter or a cosmological constant~\cite{Percacci:2002ie, 
Percacci:2003jz, Christiansen:2017cxa, Eichhorn:2018, Eichhorn:2017, Hamada:2017, Litim:2018}. 
The existence of an interacting UV fixed point implies there is an asymptotically safe quantum theory of gravity.
The effective low energy theory resulting from this is determined by solving an exact functional renormalization 
group (RG) flow equation and depends on a momentum shell parameter $k$\,. 
%
%
%
 One way to include this 
scale-dependence is to  project this RG flow from the infinite dimensional space of all functionals 
into a two dimensional subspace involving only $\sqrt{g}$ and $\sqrt{g}R$ 
(Einstein-Hilbert truncation) and write Einstein's equation in terms of the ``running" Newton's constant  
and the ``running" cosmological constant which are both dependent on the energy scale of the 
problem. In the context of cosmology, this scale may be taken to be a function only of the cosmological time,
so the running constants become time-dependent.

The Friedmann-Lema\^itre-Robertson-Walker (FLRW) model of cosmology was studied in~\cite{Bonanno:2001xi} 
using this ``RG-improved" Einstein's equation, leading to coupled ordinary differential equations for 
the scale factor $a(t)$\,, the density $\rho(t)$\,, Newton's constant $G(t)$\,, cosmological constant 
$\Lambda(t)$\,, and a ``cutoff function'' $R(0)$ which suppresses modes with momenta below the cutoff $k$ 
inside loops. In this paper we investigate Bianchi I models of cosmology using RG-improved Einstein's equation,
to check if this scheme introduces additional conditions under which the Bianchi I anisotropic cosmology approaches the FLRW universe. 
We have looked at the late time behaviour of the Bianchi-I universe for three different kinds of matter, namely, dust, radiation, and stiff matter. From the consistency conditions obtained from the renormalization group improved Einstein equations, we have found that in case of radiation, the Bianchi-I cosmological solution flows to FLRW universe at late times. However, for dust and stiff matter, the Bianchi-I universe does not necessarily flow to the FLRW universe. Further, in the case of stiff matter, we can have a Kasner type solution with some directions expanding and others contracting. This feature is not present in the radiation and dust scenarios.  

The organization of this paper is as follows. In Sec.~\ref{flow} we briefly review how the effective 
average action for gravity leads to the flow equations for the scale-dependent Newton's constant $G(k)$
and the cosmological constant $\Lambda(k)$\,. In Sec.~\ref{Bianchi} we use these to write the RG improved
Einstein's equation for the scale factors. We conclude with a discussion of our results. 

\section{Flow of $G$ and $\Lambda$}\label{flow}
In this section we briefly review the effective average action formalism for Euclidean quantum gravity in $d$ dimensions developed in \cite{Reuter:1996cp}. The analysis is based on an (Euclidean) ``effective average action'' $\Gamma_{k} [g_{\mu \nu}]$ defined such that it correctly describes all gravitational phenomena, including the effect of all loops, at a momentum scale $k$\,. Even though the quantum effective action contains all the information about the quantum theory, it turns out that it is more convenient to work with an alternative functional called the effective average action~\cite{Wetterich:1992yh,Reuter:1993kw,Reuter:1996cp,Reuter:2019book}, which is calculated like the effective action but with an infrared cutoff at the scale $k$\,. Modes with $p^2 < k^2$ are excluded while those with $p^2 > k^2$ are integrated out in the usual way. The classical action $S$ corresponds to ignoring all quantum modes, while the usual effective action $\Gamma$ corresponds to removing the IR cutoff, so $\Gamma_k$ interpolates between $S = \Gamma_{k\to \infty}$ and $\Gamma = \Gamma_{k=0}$\,. Then as a function of $k$ this $\Gamma_{k}$ describes a trajectory which satisfies a renormalization group flow equation.

The infinite dimensional space of all action functionals is then projected on the 2-dimensional subspace spanned by the functions $\sqrt{g}$ and $\sqrt{g}R$ to obtain solutions to the RG equation. For this choice of truncation in the background metric formalism, we need to consider effective actions only of the form
\begin{eqnarray}
\Gamma_{k}[g,\bar{g}]=\left(16\pi G(k)\right)^{-1} \int d^{d}x \sqrt{g}\left\lbrace -R(g)+2{\Lambda}(k)\right\rbrace + S_{gf}[g,\bar{g}]\,,
\label{EA}
\end{eqnarray}
where $\bar{g}_{\mu \nu}$ is a background metric and $S_{gf}[g,\bar{g}]$ is a classical background gauge fixing term. 
It is possible to truncate so as to include higher derivative invariants. However, for a three-dimensional subspace
it is known that the flow is essentially two dimensional close to the fixed point. Further, the projected 
2-dimensional flow gets nicely approximated by the Einstein-Hilbert flow~\cite{Lauscher:2002sq}. 

The flow equation then reads
\begin{align}
\partial_{t}\Gamma_{k}[g,\bar{g}]&=\frac{1}{2}\Tr\left[\left(\kappa^{-2}\Gamma_{k}^{(2)}[g,\bar{g}]+\mathcal{R}_{k}^{grav}[\bar{g}]\right)^{-1}\partial_{t}\mathcal{R}_{k}^{grav}[\bar{g}] \right]
\nonumber \\ &\qquad\qquad - \Tr\left[\left(-M[g,\bar{g}]+\mathcal{R}_{k}^{gh}[\bar{g}]\right)^{-1}\partial_{t}\mathcal{R}_{k}^{gh}[\bar{g}] \right]\,,
\label{FE} 
\end{align}
where we have defined $t\equiv \ln k$ and written
$\Gamma_{k}^{(2)}[g,\bar{g}]$ for the Hessian of $\Gamma_{k}[g,\bar{g}]$ with respect to $g_{\mu \nu}$. We have also defined $\kappa = ({32\pi \bar{G}})^{-\frac{1}{2}}$\,, where $\bar{G}$ is the value of $G(k)$ as $k\to \infty$\,.
Here $M$ is the Faddeev-Popov ghost operator, while $\mathcal{R}_{k}^{grav}[\bar{g}]$ and $\mathcal{R}_{k}^{gh}[\bar{g}]$ are the IR cutoff functions for gravity and the ghost operator, respectively. 

We will take both of these to be of the form $\mathcal{R}_{k}(p^2) \propto k^2 R^{(0)}\left(p^2/k^2\right)$ where the function $R^{(0)}(z)$ is smooth and satisfies the conditions $R^{(0)}(0)=1$ and $R^{(0)}(z)\rightarrow 0$ for $z \rightarrow \infty$\,, but is otherwise arbitrary. In the calculation for $\Gamma_k$\,, the $p^2$ is replaced by the kinetic operator for gravitons or ghosts. Following~\cite{Bonanno:2000ep,Bonanno:2001xi,Reuter:1996cp} we will take $R^{(0)}(z)$ to be of the form 
\begin{eqnarray}
R^{(0)}(z)=z\left[\exp(z)-1\right]^{-1} \,.
\label{IR}
\end{eqnarray}
We note that other choices for the regulator function are possible~\cite{D.litim:2001, D.litim:2000plb, Reuter:2001ag}. However, as we shall see below, the choice of different regulator functions do not qualitatively change the results.
%
%
Inserting Eq.~(\ref{EA}) into the flow equation Eq.~(\ref{FE}) gives a coupled system of equations for $\tilde{g}(k)\equiv k^{2} G(k)$ and $\lambda (k) \equiv \Lambda(k)/k^2$\, in $d=4$ dimensions
\begin{align}
k \partial_{k}\tilde{g} &=(2+\eta_{N})\tilde{g}  \label{ng} \\ k\partial_{k}\lambda &=-(2-\eta_{N})\lambda+\frac{\tilde{g} }{2\pi}\left[10\Phi_{2}^{1}(-2\lambda)-8\Phi_{2}^{1}(0) 
-5\eta_{N}\tilde{\Phi}_{2}^{1}(-2\lambda) \right]\,.
\label{nlambda}
\end{align}
%
We can think of these as individual flow equations for $\tilde{g}$ and $\lambda$\,,
with
\begin{eqnarray}
\eta_{N}(\tilde{g} ,\lambda)=\frac{\tilde{g}  B_{1}(\lambda)}{1-\tilde{g}  B_{2}(\lambda)}
\label{anomalous}
\end{eqnarray}
being the anomalous dimension of the operator $\sqrt{g}R$\,, where the functions $B_{1}(\lambda)$ and $B_{2}(\lambda)$ are given by
\begin{align}
B_{1}(\lambda)&\equiv -\frac{1}{3\pi}\left[18\Phi_{2}^{2}(-2\lambda)-5\Phi_{1}^{1}(-2\lambda)+4\Phi_{1}^{1}(0)+6\Phi_{2}^{2}(0) \right]\,,\label{B1} \\ 
B_{2}(\lambda) &\equiv \frac{1}{6\pi}\left[18\tilde{\Phi}_{2}^{2}(-2\lambda)-5\tilde{\Phi}_{1}^{1}(-2\lambda)\right]\,. 
\label{B2}
\end{align}
%
The functions $\Phi_{n}^{p}(w)$ and $\tilde{\Phi}_{n}^{p}(w)$ appearing in these expressions are given by
\begin{eqnarray}
\Phi_{n}^{p}(w)&=&\frac{1}{\Gamma(n)}\int_{0}^{\infty}dz z^{n-1}\frac{R^{(0)}(z)-z{R^{(0)}}'(z)}{\left[z+R^{(0)}(z)+w \right]^{p} }\,,
\label{phi1} \\
\tilde{\Phi}_{n}^{p}(w)&=&\frac{1}{\Gamma(n)}\int_{0}^{\infty}dz z^{n-1}\frac{R^{(0)}(z)}{\left[z+R^{(0)}(z)+w \right]^{p} }\,.
\label{phi2}
\end{eqnarray}
%
%
%
Recasting Eqs.~(\ref{ng}, \ref{nlambda}) in terms of $G(k)$, $\Lambda(k)$, we finally get
\begin{align}
k \partial_{k}G(k) &=\eta_{N}G(k) \label{nG} \\ k\partial_{k}\Lambda(k) &=\eta_{N}\Lambda(k)+\frac{1}{2\pi}k^{4}G(k)\left[10\Phi_{2}^{1}(-2\Lambda(k)/k^2)-8\Phi_{2}^{1}(0) 
-5\eta_{N}\tilde{\Phi}_{2}^{1}(-2\Lambda(k)/k^2) \right]\,. 
\label{nLambda}
\end{align}
%
%
%
Using the expressions for $B_1$ and $B_2$, we can expand the anomalous dimension of the operator $\sqrt{g}R$ for $d=4$\, in powers of $k^2$\,, 
\begin{eqnarray}
\eta_{N} =k^{2} G(k) B_{1}(\Lambda(k)/k^2)\left[ 1+k^2 G(k) B_{2}(\Lambda(k)/k^2)+k^4 G^{2}(k)B^{2}_{2}(\Lambda(k)/k^2)+\cdots\right]
\,.\qquad
\label{Anomalous}
\end{eqnarray}
%
%
Eqs.~(\ref{nG}) and~(\ref{nLambda}) cannot be solved exactly. We use an iterative procedure to find the expressions for $\Lambda$ and $G$ at small $k$\,, starting with $\Lambda=0$ and $\eta_N = 0$\,. 
Then both the functions $\Phi^p_n(\Lambda/k^2)$ and $\tilde{\Phi}^p_n(\Lambda/k^2)$ are even functions of $k$ and vanish for $k\to 0$ if $p\geq 1\,.$ 
It follows that both the functions $B_1(\lambda)$ and $ B_2(\lambda) $ and thus also $\eta_N$ are even functions of $k$ at this order of iteration. 
Looking at the equations we see that it follows easily from the iterative procedure 
that both $\Lambda(k)$ and $G(k)$ can be written as power series of only even powers of $k$\,,
\begin{align}
G(k) &= G _{0} \left [ 1-\omega  G_{0} k^{2} + \omega_{1}G^{2}_{0}k^{4}+\mathcal{O}(G^{3}_{0}k^{6}) \right ]\, \label{G} \\ 
\Lambda(k) & =  \Lambda_{0} + G_{0} k^{4} \left[\nu  +\nu _{1}G_{0}k^{2}+\mathcal{O}(G^{2}_{0}k^{4})\right]\,.
\label{flamda}
\end{align}
%
The above expressions look identical to the linearized group flow of $\lambda$ and $\tilde{g} $ near the trivial fixed point $\lambda=0$\,, $\tilde{g} = 0$\,~\cite{Reuter:2001ag}. If the couplings are on a generic flow near the trivial fixed point, we will not find a sensible result, as the $\beta$-function hits a singularity at $\lambda(k) = \frac{1}{2}$ at a non-zero value of $k$ for $\Lambda _{0}>0\,,$ with $\eta_N$ diverging at that point. However, we will see later in this paper that the consistency conditions arising from the dynamics of Bianchi type-I cosmology fixes $\Lambda_{0}=0\,,$ which puts the couplings on a trajectory which hits the trivial fixed point, avoiding the singularity. Hence there is no obstruction to taking the limit $k\to 0$\,. 

In any case, the expressions in Eq.~(\ref{G}) and Eq.~(\ref{flamda})  represent the ``quantum corrected" $G$ and $\Lambda$\,. The constants  $\omega$, $\nu,~\omega_{1}$ and $\nu_{1}$ can be calculated by inserting these expansions into Eq.~(\ref{nG}) and Eq.~(\ref{nLambda})\,,
\begin{align}
\omega &= -\frac{1}{2}B_{1}(0)=\frac{1}{6\pi} \left[24 \Phi^{2}_{2}(0)-\Phi^{1}_{1}(0)\right]=\frac{4}{\pi}\left(1-\frac{\pi^{2}}{144}\right)\,, \label{omega} \\ 
\nu &= \frac{1}{4\pi}\Phi^{1}_{2}(0)\,, \label{nu} \\ 
\omega_{1} &= \omega^{2}-\frac{B_{2}(0)}{2}\omega -\frac{13\nu}{6\pi}= \omega^{2}-\frac{\omega}{3\pi}-\frac{13\nu}{6\pi}\,, \label{omega1} \\ 
\nu_{1} &= -\omega \nu+\frac{5\omega }{6\pi}\tilde{\Phi}_{2}^{1}(0)+\frac{5\nu }{3\pi}\Phi_{2}^{2}(0)= -\omega\nu+\frac{5\omega}{6\pi}+\frac{5\nu }{3\pi} \,,
\label{nu1}
\end{align} 
with $B_{2}(0)=\frac{2}{3\pi}$, $\tilde{\Phi}_{2}^{2}(0)=1$ and $\Phi_{2}^{2}(0)=1$\,. We remark that these expressions are accurate only if we 
choose $\Lambda_0 = 0\,,$ as was noted in~\cite{Reuter:1996cp,Bonanno:2001xi}. We also remark that changing the regulator function $R^{(0)}$ does not change the form of the power series expansions of $G$ and $\Lambda$\,, but will modify the values of the constants above. 
%

\section{Bianchi I universe with running $G$ and $\Lambda$}\label{Bianchi}
We are interested in the anisotropic cosmology described by Bianchi I metric  
\begin{eqnarray}
ds^2=-dt^{2}+a^{2}(t) dx^{2}+b^{2}(t) dy^{2}+c^{2}(t) dz^{2}\,.
\label{bianchi1}
\end{eqnarray}
For time varying $G$ and $\Lambda$\, this cosmological model has been studied in the presence of a perfect fluid~\cite{Beesham:1994ni, kalligas:1995, singh:2008as, pradhan:2013as}. 
Time variation of $G$ and $\Lambda$ have also been considered in flat FLRW cosmological models~\cite{kalligas:1992}.

The energy-momentum tensor of a perfect fluid represents the cosmic matter. This is of the form 
\begin{eqnarray}
T_{\mu \nu}=(p+\rho)v_{\mu}v_{\nu}+p g_{\mu \nu}
\label{fluid-emtensor}
\end{eqnarray}
where $p$ is the pressure, $\rho$ is the energy density and $v_{\mu}$ is the four velocity of the fluid which satisfies the relation $v^{\mu}v_{\mu}=-1$.
For cosmology we will need to consider a scale defined by the cosmological time $t$, so we first write Einstein's field equations of general relativity with time-varying Newton's gravitational constant $G$ and cosmological constant $\Lambda$ 
\begin{eqnarray}
R_{\mu \nu}-\frac{1}{2}Rg_{\mu \nu}=-8\pi G(t)T_{\mu \nu}+\Lambda(t)g_{\mu \nu}~.
\label{EE}
\end{eqnarray}
Here $G(t)$ and $\Lambda(t)$ are related to the scale dependent $G(k)$ and $\Lambda(k)$ and take into account the leading quantum corrections coming from the renormalization group flow.

For the metric Eq.~(\ref{bianchi1}) together with the energy-momentum tensor Eq.~(\ref{fluid-emtensor}), we obtain the usual equations for the scale factors
\begin{align}
-\frac{\ddot{b}}{b}-\frac{\ddot{c}}{c}-\frac{\dot{b}\dot{c}}{bc} &= 8\pi Gp-\Lambda  \label{4} \\ -\frac{\ddot{a}}{a}-\frac{\ddot{c}}{c}-\frac{\dot{a}\dot{c}}{ac} &= 8\pi Gp-\Lambda  \label{5} \\ -\frac{\ddot{a}}{a}-\frac{\ddot{b}}{b}-\frac{\dot{a}\dot{b}}{ab} &= 8\pi Gp-\Lambda  \label{6} \\\frac{\dot{a}\dot{b}}{ab}+\frac{\dot{b}\dot{c}}{bc}+\frac{\dot{a}\dot{c}}{ac} &= 8\pi G \rho+\Lambda\,.
\label{7}
\end{align}
Further, the covariant conservation of the energy-momentum tensor yields
\begin{eqnarray}
\dot{\rho}+(p+\rho)\left ( \frac{\dot{a}}{a}+\frac{\dot{b}}{b}+\frac{\dot{c}}{c} \right )=0\,.
\label{8}
\end{eqnarray} 
Now since the Einstein tensor is covariantly conserved, the right hand side of Eq.~(\ref{EE}) must also be covariantly conserved. This leads to the consistency equation
\begin{eqnarray}
8\pi \rho \dot{G}+ \dot{\Lambda}=0\,,
\label{9}
\end{eqnarray}
where the dot denotes time derivative.%

Two of the equations for scale factors can be combined to produce
\begin{equation}
\frac{d}{dt}\left[  \ln \left(\frac{\dot{a}}{a}-\frac{\dot{b}}{b} \right)\right]+\left(\frac{\dot{a}}{a}+\frac{\dot{b}}{b}+\frac{\dot{c}}{c} \right)=0\,,
\label{09}
\end{equation}
integrating which we get
\begin{eqnarray}
\frac{\dot{a}}{a}-\frac{\dot{b}}{b}=\frac{k_{1}}{\mathcal{R}^3(t)}\,,
\label{10}
\end{eqnarray}
where $k_{1}$ is a constant of integration and $\mathcal{R}^{3}(t)=abc$\,.
A similar calculation using the other pairs yields
\begin{eqnarray}
\frac{\dot{b}}{b}-\frac{\dot{c}}{c}=\frac{k_{2}}{\mathcal{R}^3(t)} \label{11} \\ \frac{\dot{c}}{c}-\frac{\dot{a}}{a}=\frac{k_{3}}{\mathcal{R}^3(t)}
\label{011}
\end{eqnarray}
where $k_{2}$ and $k_{3}$ are integration constants, satisfying $k_{1}+k_{2}+k_{3}=0$\,. Let us rename for convenience the integration constants as $l$, $l\beta$ and $-l(1+\beta)$. 
Integrating these equations we get
\begin{align}
a(t) &= m_{1} \mathcal{R}(t) \exp \left [ \frac{l(2+\beta)}{3} \int \frac{dt}{\mathcal{R}^{3}(t)}\right ] \label{12} \\ 
b(t) &= m_{2}\mathcal{R}(t) \exp \left [ \frac{l(\beta-1)}{3} \int \frac{dt}{\mathcal{R}^{3}(t)}\right ] \label{13} \\ 
c(t) &= m_{3} \mathcal{R}(t) \exp \left [- \frac{l(1+2\beta)}{3} \int \frac{dt}{\mathcal{R}^{3}(t)}\right ]\,,
\label{14}
\end{align}
where $m_{1},m_{2},m_{3}$ are arbitrary constants of integration satisfying $m_{1}m_{2}m_{3}=1$\,. 

In this setup, we wish to consider the late time effect of quantum gravity. Let us use the long distance perturbative series expansion of $G(k)$ and $\Lambda(k)$ of Eq.~(\ref{G}) and Eq.~(\ref{flamda}), suitably converted to a time-varying form.
%
%
%
%
The identification of the infrared cutoff for momentum scale $k$ involves expressing $k$ in terms of all scales that are relevant to the problem under consideration. In the case of the FLRW universe, homogeneity and isotropy of spacetime imply that $k$ is a function of the cosmological time only. Hence the constants $G$ and $\Lambda$ take the form 
\begin{eqnarray}
G(t)\equiv G(k=k(t)),\qquad \Lambda(t) \equiv \Lambda(k=k(t))~.
\label{17}
\end{eqnarray}
For the anisotropic Bianchi type-I spacetime, we still have homogeneity so that all scale factors are functions of cosmological time only. Let us then consider the theory at a momentum scale set by the cosmological time, $k\equiv k(t)$ in this case also. 
%


As pointed out in~\cite{Bonanno:2001xi}, there are two natural choices of scale in an FLRW universe that could relate $t$ to $k$. One is where $k \sim t^{-1}$, which is to say that the theory is cut off at a wavelength determined by how far signals can have traveled during the lifetime of the universe, disregarding the expansion of the universe.
The other choice is to include some effect of expansion by choosing $k \sim {\mathcal{ R}}^{-1}$\,. In this case, the equations have no consistent solution with this choice for the FLRW universe with ordinary matter.  For exotic matter there is a consistent solution for the choice of $k\propto {\mathcal{ R}}(t)^{-1}$\,.  As we will see below, for a Bianchi I universe we need to include higher order terms in $t^{-1}$. Then the simplest such behaviour at late times is 
\begin{equation}
k=\sum_{n} \frac{\xi_{n}}{t^n}\,.
\label{0017}
\end{equation}
We also note that a choice of different cutoff scales in different directions does not seem practicable.

For our analysis, we will keep terms up to $n=3$ in Eq.~(\ref{0017}) because we are interested in the behaviour of $G$ and $\Lambda$ up to  $\mathcal{O}\left(\frac{t_{Pl}^{4}}{t^{4}} \right)$, so we will employ the cutoff
\begin{eqnarray}
k=\frac{\xi}{t}+\frac{\sigma}{t^{2}}+\frac{\delta}{t^{3}}\,.
\label{017}
\end{eqnarray}
Inserting this expression for $k$ into the series for $G(k)$ and $\Lambda(k)$\,, we obtain the time dependent Newton's gravitational constant and cosmological constant in the perturbative or low energy regime, 
\begin{align}
G(t) &= G _{0}\left [ 1-\frac{\tilde{\omega}G _{0}}{t^{2}}\left (1+\frac{2\tilde{\sigma }}{t}
+ \frac{2\tilde{\delta }}{t^{2}}+\frac{\tilde{\sigma }^{2}}{t^{2}} \right )+ \frac{\tilde{\omega_{1}} G _{0}^{2}}{t^{4}}+\mathcal{O} \left(\frac{t_{Pl}^{6}}{t^{6}} \right ) \right ]\,, \label{18} \\ \Lambda(t) &= \Lambda _{0}+ \frac{G_{0}}{t^{4}}\left[\tilde{\nu }\left (1+\frac{4\tilde{\sigma }}{t}+ \frac{4\tilde{\delta }}{t^{2}}+\frac{6\tilde{\sigma }^{2}}{t^{2}} \right ) +\frac{\tilde{\nu _{1}}G_{0}}{t^{2}}+\mathcal{O} \left(\frac{t_{Pl}^{4}}{t^{4}} \right ) \right]\,,
\label{19}
\end{align}
%
%
where we have defined $\tilde{\omega} \equiv \omega \xi^{2},~\tilde{\omega_{1}} \equiv \omega_{1} \xi^{4},~\tilde{\nu} \equiv \nu \xi^{4}$, $\tilde{\nu_{1}} \equiv \nu_{1} \xi^6$, $\tilde{\sigma} \equiv \frac{\sigma}{\xi}$ and $\tilde{\delta} \equiv \frac{\delta}{\xi}$ for convenience. 

To proceed further we now assume, in line with the arguments of~\cite{Bonanno:2001xi}, that renormalization effects coming from the matter sector are small compared to those of pure quantum gravity. Then the equation of state relating the pressure $p$ and the energy density $\rho$ is linear,
\begin{equation}
p(t)=\Omega \rho(t)\,,
\label{20}
\end{equation}
where $\Omega$ is a constant. 
We next substitute the equation of state into the energy-momentum conservation law  Eq.~(\ref{8}) and integrate it. This gives 
\begin{equation}
\rho \left [ \mathcal{R}(t) \right ]^{3(1+\Omega)}=\frac{\mathcal{M}}{8 \pi}\,,
\label{22}
\end{equation}
where $\mathcal{M}$ is an integration constant.
On the other hand, using Eq.~(\ref{9}) we can express the energy density $\rho (t)$ in the form
\begin{equation}
\rho =-\frac{1}{8\pi}\frac{\dot{\Lambda }}{\dot{G}}\,.
\label{23}
\end{equation}
Combining Eq.~(\ref{22}) and Eq.~(\ref{23}), we obtain 
\begin{equation}
\mathcal{R}(t)=\left [ -\frac{\mathcal{M}\dot{G}}{\dot{\Lambda }} \right ]^{\frac{1}{3+3\Omega }}\,.
\label{24}
\end{equation}
The time derivatives of $G(t)$ and $\Lambda(t)$ can be calculated from their expressions above and using them in the expressions for the energy density $\rho$ and the average scale factor ${\mathcal{ R}}$, we find
%
%
 %
\begin{align}
\rho(t)&= \frac{1}{4\pi}\left(\frac{\tilde{\nu}}{\tilde{\omega }}\right)\frac{1}{G_{0}t^{2}}\left \{ 1+\frac{2\tilde{\sigma }}{t}+ \frac{2\tilde{\delta }}{t^{2}}+\frac{\tilde{\sigma }^{2}}{t^{2}}+\left(\frac{2\tilde{\omega_{1}}}{\tilde{\omega}}+\frac{3\tilde{\nu_{1}}}{2\tilde{\nu}} \right)\frac{G_{0}}{t^{2}}+\mathcal{O}\left ( \frac{G_{0}^{2}}{t^{4}} \right)\right \}\,,
 \label{26} 
\\
\mathcal{ R}(t)&=\left [ \frac{\mathcal{M} G_{0}}{2}\left ( \frac{\tilde{\omega}}{\tilde{\nu}} \right )\right ]^{\frac{1}{(3+3\Omega)}} t^{\frac{2}{(3+3\Omega)}} \left \{ 1-\frac{1}{(3+3\Omega)}\left (\frac{2\tilde{\sigma }}{t}+ \frac{2\tilde{\delta }}{t^{2}}-\frac{(3\Omega+5)}{(3+3\Omega)}\frac{\tilde{\sigma }^{2}}{t^{2}}+\right. \right. \nonumber \\ & \left. \left. \qquad\qquad\qquad\qquad\qquad\qquad\qquad\qquad\qquad \left(\frac{2\tilde{\omega_{1}}}{\tilde{\omega}}+\frac{3\tilde{\nu_{1}}}{2\tilde{\nu}} \right)\frac{G_{0}}{t^{2}}\right )+\mathcal{O}\left ( \frac{G_{0}^{2}}{t^{4}} \right)\right \}
. 
\label{27}
\end{align}
We now discuss three cases of cosmic matter separately: $i)$ dust, for which $\Omega=0$\,; $ii)$ radiation,  $\Omega=\frac{1}{3}$\,; and $iii)$ stiff fluid, $\Omega=1$\,.  The first two cases have something in common, as we will see now. For $\Omega\neq 1$ and defining $\alpha = \left [ \frac{\mathcal{M} G_{0}}{2}\left ( \frac{\tilde{\omega}}{\tilde{\nu}} \right )\right ]^{-\frac{1}{1+\Omega}}$, we can integrate the expressions for the scale factors to find 
\begin{align}
a(t) &= m_{1} \mathcal{R}(t)\exp\left [ \frac{l(2+\beta)\alpha }{3}\mathcal{N}(t) \right]\,, 
 \notag \\
b(t) &= m_{2} \mathcal{R}(t) \exp\left [ \frac{l(\beta-1)\alpha }{3}\mathcal{N}(t) \right]\,,
 \notag \\
c(t) &= m_{3} \mathcal{R}(t) \exp\left [- \frac{l(1+2\beta)\alpha }{3}\mathcal{N}(t)\right]\,,
\label{31}
\end{align}
where we have written 
\begin{align}
\mathcal{N}(t)&=\int \frac{dt}{\mathcal{R}^{3}(t)} \nonumber \\ &  = \frac{(\Omega +1)}{(\Omega -1)} t^{\frac{(\Omega -1)}{(\Omega +1)}}-\tilde{\sigma }t^{-\frac{2}{(\Omega +1)}}-\frac{1}{(\Omega+3)}\left (2\tilde{\delta }-\frac{(\Omega-1)\tilde{\sigma }^{2}}{(\Omega+1)}+\left(\frac{2\tilde{\omega_{1}}}{\tilde{\omega}}+\frac{3\tilde{\nu_{1}}}{2\tilde{\nu}} \right)G_{0}\right ) t^{-\frac{(\Omega +3)}{(\Omega +1)}}\,,
\label{031}
\end{align}
and we have neglected higher order terms. 
%
%
From these equations for the scale factors, we can compute the directional Hubble parameters,
\begin{eqnarray}
\frac{\dot{a}}{a}&= H(t) + \frac{l(2+\beta)\alpha }{3}H_{1}(t) \,,
\notag\\ 
\frac{\dot{b}}{b} &=  H(t) + \frac{l(\beta-1)\alpha }{3}H_{1}(t) \,,
\notag\\ 
\frac{\dot{c}}{c} &=  H(t) - \frac{l(1+2\beta)\alpha }{3}H_{1}(t)\,. 
\label{34} 
\end{eqnarray}
%
%
Here the ``average Hubble parameter'' $H(t)$ includes isotropic quantum corrections,
\begin{equation}
H(t)=\frac{\dot{\mathcal{R}}}{\mathcal{R}}=\frac{2}{(3+3\Omega )}\frac{1}{t}\left [ 1+\frac{\tilde{\sigma }}{t}+\left ( 2\tilde{\delta}-\tilde{\sigma }^{2}+\left(\frac{2\tilde{\omega_{1}}}{\tilde{\omega}}+\frac{3\tilde{\nu_{1}}}{2\tilde{\nu}} \right)G_{0} \right )\frac{1}{t^{2}} 
+\mathcal{O}\left (\frac{t_{Pl}^{3}}{t^{3}}  \right ) \right ]\,,
\label{Hubble-avg}
\end{equation}
while the effects of anisotropy are included in the coefficients of $H_1(t)$\,, which also includes quantum corrections,
\begin{equation}
H_{1}(t)=t^{-\frac{2}{1+\Omega }}+ \frac{2\tilde{\sigma }}{(1+\Omega)}t^{-\frac{(\Omega+3)}{(1+\Omega) }}+\frac{1}{(\Omega+1)}\left (2\tilde{\delta }-\frac{(\Omega-1)\tilde{\sigma }^{2}}{(\Omega+1)}+\left(\frac{2\tilde{\omega_{1}}}{\tilde{\omega}}+\frac{3\tilde{\nu_{1}}}{2\tilde{\nu}} \right)G_{0}\right ) t^{-\frac{2(2+\Omega )}{1+\Omega }}\,.
\label{Hubble-corr}
\end{equation} 
%

We now run a consistency check on the solutions for the scale factors,  by putting these solutions into Eq.~(\ref{7}). Keeping up to $\mathcal{O}\left(\frac{t_{Pl}}{t}\right)^4$\,, we find
%
%
%
%
\begin{eqnarray}
\notag
& 3\left( \frac{\dot{\mathcal{R}}}{\mathcal{R}}\right)^{2}-\frac{\alpha^{2} l^2( \beta^{2}+\beta+1 )}{3}t^{-\frac{4}{1+\Omega }}\left \{1+\frac{4\tilde{\sigma}}{(1+\Omega )}\frac{1}{t}+\frac{2}{(1+\Omega )}\left (2\tilde{\delta }+\frac{(3-\Omega )}{(1+\Omega )}\tilde{\sigma}^{2}+\left(\frac{2\tilde{\omega_{1}}}{\tilde{\omega}}+\frac{3\tilde{\nu_{1}}}{2\tilde{\nu}} \right)G_{0}\right ) \frac{1}{t^{2}}\right \} 
\\ & \qquad=\Lambda _{0} +2(\frac{\tilde{\nu }}{\tilde{\omega }})\frac{1}{t^{2}}+4(\frac{\tilde{\nu }}{\tilde{\omega }})\frac{\tilde{\sigma }}{t^{3}}-\tilde{\nu}\frac{G_{0}}{t^4}+2(\frac{\tilde{\nu }}{\tilde{\omega }})\left ( 2\tilde{\delta }+\tilde{\sigma}^{2}+\left(\frac{2\tilde{\omega_{1}}}{\tilde{\omega}}+\frac{3\tilde{\nu_{1}}}{2\tilde{\nu}} \right)G_{0} \right )\frac{1}{t^4} \,.
\label{consistency}
\end{eqnarray}
%
The above equation is the consistency relation which the scale factor must satisfy for all $\Omega\neq1$\,. Let us now see the consequences of the above relation for $\Omega=0$ (dust) and $\Omega=\frac{1}{3}$ (radiation)\,.
%
\subsection{$\Omega=0$}
Comparing the coefficients of different powers of $t$ on both sides of Eq.~(\ref{consistency}) we get the following consistency conditions 
\begin{align}
\Lambda_{0} &= 0 \,,
\label{37}\\ 
\frac{\tilde{\omega}}{\tilde{\nu}} &= \frac{3}{2}\,, 
\label{38}\\
\frac{4}{3}\left [4\tilde{\delta}-\tilde{\sigma}^{2}
+2\left(\frac{2\tilde{\omega_{1}}}{\tilde{\omega}}+\frac{3\tilde{\nu_{1}}}{2\tilde{\nu}} \right)G_{0} \right ]& -\frac{l^2(\beta^{2}+\beta+1)\alpha^{2}}{3}=-\tilde{\nu}G_{0}+ \nonumber \\& 2\left ( \frac{\tilde{\nu }}{\tilde{\omega}} \right )\left [2\tilde{\delta}+\tilde{\sigma}^{2}+\left(\frac{2\tilde{\omega_{1}}}{\tilde{\omega}}+\frac{3\tilde{\nu_{1}}}{2\tilde{\nu}} \right)G_{0}\right ] \,,
\label{39}
\end{align}
%
with $\alpha = \left [ \frac{\mathcal{M} G_{0}}{2}\left ( \frac{\tilde{\omega}}{\tilde{\nu}} \right )\right ]^{-1}$\, for $\Omega=0$.

Combining Eq.~(\ref{38}) and Eq.~(\ref{39}), we get a consistency condition valid up to $\mathcal{O}\left(\frac{1}{t^4}\right)$ in our calculations,
\begin{equation}
\frac{8}{3}\tilde{\sigma }^{2}-\frac{8}{3}\tilde{\delta} = \tilde{\nu}G_{0}+\frac{4G_{0}}{3}\left(\frac{2\tilde{\omega_{1}}}{\tilde{\omega}}+\frac{3\tilde{\nu_{1}}}{2\tilde{\nu}} \right)-\frac{l^2 \alpha ^{2}\left(\beta^{2}+\beta+1 \right)}{3} \,.
\label{039}
\end{equation}
Note that if we had not included the $\mathcal{O}\left(\frac{1}{t^4}\right)$ terms in Eq.~(\ref{consistency}), 
we would have obtained conditions corresponding to FLRW cosmology, which were found in~\cite{Bonanno:2001xi}. 
If we keep terms up to $\mathcal{O}\left(\frac{1}{t^4}\right)$ and compare coefficients, we can immediately conclude that for $l=0$ we will regain the FLRW universe. Thus we see that for $\Omega=0$, the anisotropic Bianchi-I cosmology does not necessarily flow to the FLRW solution when quantum corrections are included.   
%
%
%
%
%

\subsection{$\Omega=\frac{1}{3}$}
In this case,  by comparing inverse powers of $t$ in the consistency condition Eq.~(\ref{consistency}), we again find $\Lambda_{0} = 0$\,, and 
\begin{align}
\frac{\tilde{\omega}}{\tilde{\nu}} &= \frac{8}{3}\,, \label{40} \\ 4\left ( \frac{\tilde{\nu }}{\tilde{\omega }}\right )\tilde{\sigma } &= \frac{3}{2}\tilde{\sigma }-\frac{l^2 \alpha ^{2}\left(\beta^{2}+\beta+1 \right)}{3} \,,
\label{41}
\end{align}
%
from which it immediately follows that
\begin{equation}
l^2 \alpha ^{2}\left(\beta^{2}+\beta+1 \right)=0\,.
\label{041}
\end{equation}
If $l\neq 0$, we must have $ \beta^{2}+\beta+1 =0 $\,. This implies that the two roots of $\beta$ are complex. 
Since the scale factors must be real, it follows that $l=0$\,. Then from the terms of order $t^{-4}$ in Eq.~(\ref{consistency}) 
we get the condition 
\begin{eqnarray}
\frac{3}{2}\left ( \tilde{\sigma}^{2}-\tilde{\delta}\right )= \tilde{\nu}G_{0}+\frac{3}{4}\left ( \frac{2\tilde{\omega_{1}}}{\tilde{\omega}}+\frac{3\tilde{\nu_{1}}}{2\tilde{\nu}} \right)G_{0}~.
\label{0041}
\end{eqnarray}
%
%
Hence we see that all the directional Hubble parameters must be equal, i.e. the universe must be FLRW, in the presence of radiation. 
It is also not difficult to see that for $0<\Omega<1$ Eq.~(\ref{041}) will always appear as a consistency condition, so $l=0$ and the 
universe becomes FLRW at late times. Thus we can conclude from the above analysis that the scale factors of anisotropic Bianchi type-I metric flow to the isotropic FLRW cosmology due to renormalization group flow of the Newton's gravitational constant $G(t)$ and the cosmological constant $\Lambda(t)$, for all  $0<\Omega<1$\,. 
%

\subsection{$\Omega=1$}
The case of $\Omega=1$, which corresponds to stiff matter, is somewhat different. First we write down the expression for the average of the scale factor $\mathcal{R}$ by setting $\Omega=1$ in Eq.~(\ref{27}). This produces
\begin{equation}
\mathcal{ R}(t)=\left [ \frac{\mathcal{M} G_{0}}{2}\left ( \frac{\tilde{\omega}}{\tilde{\nu}} \right )\right ]^{\frac{1}{6}} t^{\frac{1}{3}} \left \{ 1-\frac{1}{6}\left (\frac{2\tilde{\sigma }}{t}+ \frac{2\tilde{\delta }}{t^{2}}-\frac{4}{3}\frac{\tilde{\sigma }^{2}}{t^{2}}+\left(\frac{2\tilde{\omega_{1}}}{\tilde{\omega}}+\frac{3\tilde{\nu_{1}}}{2\tilde{\nu}} \right)\frac{G_{0}}{t^{2}}\right )+\mathcal{O}\left ( \frac{t_{Pl}^{3}}{t^{3}} \right)\right \}\,.
 \\
\label{44}
\end{equation}
The solutions for the scale factors for stiff matter are then
\begin{align}
a(t) &= m_{1} \mathcal{R}(t)\, t^{\frac{l(2+\beta)\alpha}{3}} \exp\left [- \frac{l(2+\beta)\alpha }{3}\mathcal{Q}(t)\right]\,
 \notag \\ 
b(t)&= m_{2} \mathcal{R}(t)\, t^{\frac{l(\beta-1)\alpha}{3}} \exp\left [ -\frac{l(\beta-1)\alpha}{3} \mathcal{Q}(t) \right]\,  \notag \\ 
c(t)&= m_{3} \mathcal{R}(t)\, t^{-\frac{l(1+2\beta)\alpha}{3}} \exp\left [\frac{l(1+2\beta)\alpha}{3}\mathcal{Q}(t) \right] \,,
\label{47}
\end{align}
where we now have $\alpha=\left [ \frac{\mathcal{M} G_{0}}{2}\left ( \frac{\tilde{\omega}}{\tilde{\nu}} \right )\right ]^{-\frac{1}{2}}$ for $\Omega=1$\, and
\begin{eqnarray}
\mathcal{Q}(t)=\tilde{\sigma }t^{-1}+\frac{1}{4}\left (2\tilde{\delta }+\left(\frac{2\tilde{\omega_{1}}}{\tilde{\omega}}+\frac{3\tilde{\nu_{1}}}{2\tilde{\nu}} \right)G_{0}\right ) t^{-2}~. 
\label{047}
\end{eqnarray}

The directional Hubble parameters are now computed up to $\mathcal{O} \left((\frac{t_{Pl}}{t})^3\right)$\,,
\begin{align}
\frac{\dot{a}}{a} &=  \frac{\dot{\mathcal{R}}}{\mathcal{R}}+ \frac{l(2+\beta)\alpha }{3} \bar{H}(t)\, \notag \\ 
\frac{\dot{b}}{b} &=  \frac{\dot{\mathcal{R}}}{\mathcal{R}}+ \frac{l(\beta-1)\alpha }{3}\bar{H}(t) \, \notag \\ 
\frac{\dot{c}}{c} &=  \frac{\dot{\mathcal{R}}}{\mathcal{R}}- \frac{l(1+2\beta)\alpha }{3} \bar{H}(t)\,,
\label{50} 
\end{align}
The (isotropic) average Hubble parameter is
\begin{eqnarray}
\frac{\dot{\mathcal{R}}}{\mathcal{R}}=\frac{1}{3t}\left [ 1+\frac{\tilde{\sigma }}{t}+\left ( 2\tilde{\delta}-\tilde{\sigma }^{2}+\left(\frac{2\tilde{\omega_{1}}}{\tilde{\omega}}+\frac{3\tilde{\nu_{1}}}{2\tilde{\nu}} \right)G_{0} \right )\frac{1}{t^{2}}+\mathcal{O}\left (\frac{t_{Pl}^{3}}{t^{3}}  \right ) \right ]
\label{51}
\end{eqnarray}
and we have also written, to the same order of approximation, 
\begin{eqnarray}
\bar{H}(t)=\frac{1}{t}+\frac{\tilde{\sigma}}{t^{2}}+\frac{1}{2}\left (2\tilde{\delta }+\left(\frac{2\tilde{\omega_{1}}}{\tilde{\omega}}+\frac{3\tilde{\nu_{1}}}{2\tilde{\nu}} \right)G_{0}\right )\frac{1}{t^{3}}~.
\label{051}
\end{eqnarray}
%
%

As before, we now put these solutions into Eq.~(\ref{7}) for a consistency check. The consistency condition is then found to be
\begin{align}
& 3\left( \frac{\dot{\mathcal{R}}}{\mathcal{R}}\right)^{2}-\frac{\alpha^{2} l^2( \beta^{2}+\beta+1 )}{3}\left \{\frac{1}{t^{2}}+\frac{\tilde{2\sigma}}{t^{3}}+\left (2\tilde{\delta }+\left(\frac{2\tilde{\omega_{1}}}{\tilde{\omega}}+\frac{3\tilde{\nu_{1}}}{2\tilde{\nu}} \right)G_{0}\right )\frac{1}{t^{4}}+\frac{\tilde{\sigma }^{2}}{t^{4}}\right \}  \nonumber  \\
 &\qquad=\Lambda _{0} +\frac{2\tilde{\nu }}{\tilde{\omega }t^2} +\frac{4\tilde{\nu }\tilde{\sigma}}{\tilde{\omega }t^{3}}
 -\tilde{\nu}\frac{G_{0}}{t^4}+\frac{2\tilde{\nu }}{\tilde{\omega }}\left ( 2\tilde{\delta }+\tilde{\sigma}^{2}+\left(\frac{2\tilde{\omega_{1}}}{\tilde{\omega}}+\frac{3\tilde{\nu_{1}}}{2\tilde{\nu}} \right)G_{0} \right )\frac{1}{t^4} ~. 
\label{53}
\end{align}
Calculating $\left( \frac{\dot{\mathcal{R}}}{\mathcal{R}}\right)^{2}$ from Eq.~(\ref{51}) and 
comparing the coefficients of the $t^0\,, t^{-2}\,, t^{-4}$ terms respectively, we find the equations 
\begin{align}
\Lambda_{0} &= 0\,,
 \label{54} \\
2\left ( \frac{\tilde{\nu}}{\tilde{\omega}} \right ) &= \frac{1}{3}\left( 1-\alpha^{2}l^2( \beta^{2}+\beta+1 ) \right) \,,
\label{55}\\
\frac{2}{3}\left \{ 2\tilde{\delta}-\frac{\tilde{\sigma}^{2}}{2}+\left(\frac{2\tilde{\omega}_{1}}{\tilde{\omega}}+\frac{3\tilde{\nu}_{1}}{2\tilde{\nu}}\right)G_{0} \right \} &-\frac{l^{2}\alpha^{2}(\beta^{2}+\beta+1)}{3}  \left \{ 2\tilde{\delta}+  \tilde{\sigma}^{2}+\left(\frac{2\tilde{\omega}_{1}}{\tilde{\omega}}+\frac{3\tilde{\nu}_{1}}{2\tilde{\nu}}\right)G_{0} \right \} \nonumber \\ &=-\tilde{\nu}G_{0}+ \frac{2\tilde{\nu }}{\tilde{\omega }} \left ( 2\tilde{\delta }+\tilde{\sigma}^{2}+\left(\frac{2\tilde{\omega}_{1}}{\tilde{\omega}}+\frac{3\tilde{\nu}_{1}}{2\tilde{\nu}} \right)G_{0} \right )\,. 
\label{56} 
\end{align}
Using Eq.~(\ref{55}) in Eq.~(\ref{56}), we obtain
\begin{eqnarray}
\frac{2}{3}\left ( \tilde{\sigma}^{2}-\tilde{\delta}\right )= \tilde{\nu}G_{0}+\frac{1}{3}\left ( \frac{2\tilde{\omega_{1}}}{\tilde{\omega}}+\frac{3\tilde{\nu_{1}}}{2\tilde{\nu}} \right )G_{0}~.
\label{056}
\end{eqnarray}
%
%
We note that from consistency condition in Eq.~(\ref{55}), we can write $l$ in terms of other constant $\beta$ for $\Omega=1$\,,
\begin{eqnarray}
l &=& \frac{1}{\alpha}\frac{\sqrt{1-6(\frac{\tilde{\nu}}{\tilde{\omega}})}}{\sqrt{ \left ( \beta ^{2}+\beta +1 \right )}}~\,.
\label{57}
\end{eqnarray}
As $\frac{\dot{a}}{a},\frac{\dot{b}}{b}$ and $\frac{\dot{c}}{c}$ are real, we get the following condition from the above equation 
\begin{equation}
1-6\left(\frac{\tilde{\nu}}{\tilde{\omega}}\right) \geq  0  \qquad 
\Rightarrow \qquad \xi^2 \leq  \frac{1}{6} \frac{\omega}{\nu}~.
\label{64}
\end{equation}
Note that the inequality is saturated for the FLRW case, since from Eq.~(\ref{55}) we see that $\frac{\tilde{\nu}}{\tilde{\omega}}=\frac{1}{6}$ when $l^2( \beta^{2}+\beta+1 )=0$\,, which implies that $l=0$ since $\beta$ must be real.

Finally, by putting Eq.~(\ref{57}) into Eq.~(\ref{50}) we observe that for large $\beta$, we get a Kasner type solution, i.e. there are expanding and contracting directions. For large positive $\beta$ the expanding directions would involve the scale factors $a$, $b$ and contracting direction would involve $c$. We further find that the range of  $\beta$ which result in a Kasner type solution change when we take into account the quantum gravity corrections. Likewise we get a Kasner solution for some values of negative $\beta$ as well.

\section{Conclusions}
%
In this paper, we have studied the anisotropic Bianchi-I cosmological model taking quantum gravitational effects into account. The analysis is valid for late times which correspond to the perturbative regime of the exact renormalization group flow of the effective average action for quantum gravity. We used the renormalization group improved cosmological evolution equation by including the scale dependence of Newton's constant and the cosmological constant. We have obtained the solution of $G$ and $\Lambda$ in power series of the infrared cutoff scale from the cosmological evolution equation. 

In an improvement over previous works, we have included higher powers of $1/t$ in the expression for the infrared cutoff scale $k$\,. If we had not done this, the consistency condition Eq.~(\ref{consistency}) would not hold for any value of $\Omega < 1\,.$ Indeed, because we have considered higher order terms in the expansion of $G$ and $\Lambda$ in terms of powers of the infrared cutoff, the consistency conditions would fail even for isotropic cosmologies without the additional terms in the expression for $k\,.$
%
%
From this we have found the solution of the energy density and average scale factor in an inverse power series of the cosmological time. An important point to note in this regard is that the power series expansion of the dimensionful cosmological constant $\Lambda$ makes sense only if the leading term $\Lambda_0$ vanishes, because otherwise the dimensionless $\lambda$ diverges as $k\to 0$. But if $\Lambda_0 = 0$\,, the dimensionless couplings $\tilde{g}$ and $\lambda$ flow on a trajectory directed towards the trivial fixed point $\tilde{g}=0\,, \lambda=0$ as the infrared cutoff scale goes to zero. Thus it comes as no surprise that the consistency of Einstein equations with renormalization group flow analysis of $\tilde{g}$ and $\lambda$ implies that the only allowed value of $\Lambda_0$ is zero, for all types of fluids, namely, dust, radiation and stiff matter.

Using these solutions for $G$ and $\Lambda$, we have then showed how the flow of anisotropic Bianchi-I cosmology gets affected by quantum gravitational effects for known matter like the dust, radiation and stiff matter. We have computed the scale factors from Einstein equations for dust, radiation and stiff matter for Bianchi-I metric. The consistency conditions following from Einstein equations indicate that the Bianchi-I anisotropic cosmological universe eventually evolves into a FLRW universe at late times if filled with a perfect fluid with the equation of state $p=\Omega \rho$ for $0<\Omega < 1$. This includes the case of radiation. The scale factors $a(t)\,,b(t)$ and $c(t)$ take the same form and expand in the same rate in all directions. For the $\Omega=0$ case which corresponds to dust, we find that the Bianchi-I universe does not necessarily flow to the FLRW isotropic universe.
For $\Omega=1$ which corresponds to stiff matter, we observe from the consistency conditions that the solution does not flow to the isotropic FLRW universe at late times. We also calculate a bound on the cutoff parameter $\xi$ and find that the Bianchi-I universe flows to the isotropic FLRW univese at late times if $\xi^2$ equals its maximum value, but not otherwise. We also find that there is a possibility of getting a Kasner like solution in this case.

\section*{Acknowledgments} RM would like to thank DST-INSPIRE, Govt. of India for financial support. The authors thank the anonymous referees for relevant suggestions.



\begin{thebibliography}{99}
	
\bibitem{Will:2018bme} 
C.~M.~Will,
``Theory and Experiment in Gravitational Physics,''
Cambridge University Press, London U,K, (2018).
	
\bibitem{Will:2014kxa} 
C.~M.~Will,
``The Confrontation between General Relativity and Experiment,''
Living Rev.\ Rel.\  {\bf 17}, 4 (2014)
doi:10.12942/lrr-2014-4

\bibitem{Utiyama:1956sy} 
R.~Utiyama,
``Invariant theoretical interpretation of interaction,''
Phys.\ Rev.\  {\bf 101}, 1597 (1956).
doi:10.1103/PhysRev.101.1597

\bibitem{Weinberg:1972kfs} 
S.~Weinberg,
``Gravitation and Cosmology : Principles and Applications of the General Theory of Relativity,''
John Wiley and Sons, New York (1972).

\bibitem{Hehl:1976my} 
F.~W.~Hehl, G.~D.~Kerlick and P.~Von Der Heyde,
``On a New Metric Affine Theory of Gravitation,''
Phys.\ Lett.\  {\bf 63B}, 446 (1976).
doi:10.1016/0370-2693(76)90393-2

\bibitem{Hehl:1976kj} 
F.~W.~Hehl, P.~Von Der Heyde, G.~D.~Kerlick and J.~M.~Nester,
``General Relativity with Spin and Torsion: Foundations and Prospects,''
Rev.\ Mod.\ Phys.\  {\bf 48}, 393 (1976).
doi:10.1103/RevModPhys.48.393

\bibitem{Blagojevic:2002du} 
M.~Blagojevi\'c,
``Gravitation and gauge symmetries,''
Bristol, UK: IOP (2002).

\bibitem{Blagojevic:2013xpa} 
M.~Blagojevi\'c  and F.~W.~Hehl,
``Gauge Theories of Gravitation : A Reader with Commentaries,''
World Scientific, Singapore (2013).

\bibitem{rovelli:2004} 
C.~Rovelli ,
``Quantum Gravity (Cambridge Monographs on Mathematical Physics) ,''
Cambridge University Press (2004).

\bibitem{thiemann:2004} 
T.~Thiemann,
``Lectures on Loop Quantum Gravity ,''
Lect.\ Notes Phys.\ 631:41-135,2003.
[arXiv:gr-qc/0210094].

\bibitem{wess:1992} 
J.Wess and J.~Bagger,
``Supersymmetry and Supergravity ,''
Princeton University Press (1992).

\bibitem{Bern:2009kd} 
Z.~Bern, J.~J.~Carrasco, L.~J.~Dixon, H.~Johansson and R.~Roiban,
``The Ultraviolet Behavior of N=8 Supergravity at Four Loops,''
Phys.\ Rev.\ Lett.\  {\bf 103}, 081301 (2009)
doi:10.1103/PhysRevLett.103.081301
[arXiv:0905.2326 [hep-th]].

\bibitem{polchinski:1998a} 
J.~Polchinski,
`` String Theory Vol. I: An Introduction to the Bosonic String ,''
Cambridge University Press (1998).

\bibitem{polchinski:1998b} 
J.~Polchinski,
``String Theory Vol. II: Superstring Theory and Beyond,''
Cambridge University Press (1998).

\bibitem{kiritsis:2007} 
E.~Kiritsis ,
``String Theory in a Nutshell ,''
Princeton University Press (2007).


\bibitem{donoghue:1994} 
J.~F.~Donoghue,
``General relativity as an effective field theory: The leading quantum correctionsr,''
Phys.\ Rev.\ D {\bf 50} (1994) 3874-3888.
doi:10.1103/PhysRevD.50.3874
[arXiv:gr-qc/9405057].

\bibitem{verlinde:2011} 
E.~P.~Verlinde,
``On the Origin of Gravity and the Laws of Newton,''
JHEP 1104:029,2011.
doi:10.1007/JHEP04(2011)029.
[arXiv:1001.0785v1 [hep-th]] 

\bibitem{padmanabhan:2010} 
T.~Padmanabhan,
``Thermodynamical Aspects of Gravity: New insights,''
Rep.\ Prog.\ Phys.\ {\bf 73} (2010) 046901.
doi:10.1088/0034-4885/73/4/046901
[arXiv:0911.5004v2 [gr-qc]] 


\bibitem{Weinberg:1980gg} 
S.~Weinberg,
``Ultraviolet Divergences In Quantum Theories Of Gravitation,''
in General Relativity: An Einstein Centenary Survey, S.~W.~Hawking and
W.~Israel (eds.) University Press, Cambridge (1979). 

\bibitem{Gross:1973id} 
D.~J.~Gross and F.~Wilczek,
``Ultraviolet Behavior of Nonabelian Gauge Theories,''
Phys.\ Rev.\ Lett.\  {\bf 30}, 1343 (1973).
doi:10.1103/PhysRevLett.30.1343

\bibitem{Politzer:1973fx} 
H.~D.~Politzer,
``Reliable Perturbative Results for Strong Interactions?,''
Phys.\ Rev.\ Lett.\  {\bf 30}, 1346 (1973).
doi:10.1103/PhysRevLett.30.1346

\bibitem{Reuter:1996cp}
M.~Reuter,
``Nonperturbative evolution equation for quantum gravity,''
Phys.\ Rev. \ D {\bf 57}, no. 10, 971 (1998).
doi:10.1103/PhysRevD.57.971
[arXiv:hep-th/9605030].


\bibitem{Lauscher:2001ya} 
O.~Lauscher and M.~Reuter,
``Ultraviolet fixed point and generalized flow equation of quantum gravity,''
Phys.\ Rev.\ D {\bf 65}, 025013 (2002).
doi:10.1103/PhysRevD.65.025013
[hep-th/0108040].

\bibitem{Lauscher:2002sq} 
O.~Lauscher and M.~Reuter,
``Flow equation of quantum Einstein gravity in a higher derivative truncation,''
Phys.\ Rev.\ D {\bf 66}, 025026 (2002).
doi:10.1103/PhysRevD.66.025026
[hep-th/0205062].

\bibitem{Souma:1999at} 
W.~Souma,
Prog.\ Theor.\ Phys.\  {\bf 102}, 181 (1999).
doi:10.1143/PTP.102.181
[hep-th/9907027].

\bibitem{Percacci:2005wu} 
R.~Percacci,
Phys.\ Rev.\ D {\bf 73}, 041501 (2006).
doi:10.1103/PhysRevD.73.041501
[hep-th/0511177].

\bibitem{Niedermaier:2006wt} 
M.~Niedermaier and M.~Reuter,
``The Asymptotic Safety Scenario in Quantum Gravity,''
Living Rev.\ Rel.\  {\bf 9}, 5 (2006).
doi:10.12942/lrr-2006-5

\bibitem{Niedermaier:2006ns} 
M.~Niedermaier,
``The Asymptotic safety scenario in quantum gravity: An Introduction,''
Class.\ Quant.\ Grav.\  {\bf 24}, R171 (2007).
doi:10.1088/0264-9381/24/18/R01
[gr-qc/0610018].

\bibitem{Niedermaier:2010zz} 
M.~Niedermaier,
``Gravitational fixed points and asymptotic safety from perturbation theory,''
Nucl.\ Phys.\ B {\bf 833}, 226 (2010).
doi:10.1016/j.nuclphysb.2010.01.016


\bibitem{Falls:2014tra} 
K.~Falls, D.~F.~Litim, K.~Nikolakopoulos and C.~Rahmede,
``Further evidence for asymptotic safety of quantum gravity,''
Phys.\ Rev.\ D {\bf 93}, no. 10, 104022 (2016).
doi:10.1103/PhysRevD.93.104022
[arXiv:1410.4815 [hep-th]].

\bibitem{Percacci:2002ie} 
R.~Percacci and D.~Perini,
``Constraints on matter from asymptotic safety,''
Phys.\ Rev.\ D {\bf 67}, 081503 (2003).
doi:10.1103/PhysRevD.67.081503
[hep-th/0207033].

\bibitem{Percacci:2003jz} 
R.~Percacci and D.~Perini,
``Asymptotic safety of gravity coupled to matter,''
Phys.\ Rev.\ D {\bf 68}, 044018 (2003).
doi:10.1103/PhysRevD.68.044018
[hep-th/0304222].

\bibitem{Christiansen:2017cxa} 
N.~Christiansen, D.~F.~Litim, J.~M.~Pawlowski and M.~Reichert,
``Asymptotic safety of gravity with matter,''
Phys.\ Rev.\ D {\bf 97}, no. 10, 106012 (2018).
doi:10.1103/PhysRevD.97.106012
[arXiv:1710.04669 [hep-th]].


\bibitem{Eichhorn:2018} 
A.~Eichhorn, S.~Lippoldt and V.~Skrinjar,
``Nonminimal hints for asymptotic safety,''
Phys.\ Rev.\ D {\bf 97}, 026002 (2018).
doi:10.1103/PhysRevD.97.026002
[arXiv:1710.03005v2 [hep-th]].


\bibitem{Eichhorn:2017} 
A.~Eichhorn and A.~Held,
``Viability of quantum-gravity induced ultraviolet completions for matter,''
Phys.\ Rev.\ D {\bf 96}, 086025 (2017).
doi:10.1103/PhysRevD.96.086025
[arXiv:1705.02342v2 [gr-qc]].


\bibitem{Hamada:2017} 
Y.~Hamada and M.~Yamada,
``Asymptotic safety of higher derivative quantum gravity non-minimally coupled with a matter system,''
JHEP (2017) 2017: {\bf 70}.
doi:10.1007/JHEP08(2017)070
[arXiv:1703.09033v3 [hep-th]].


\bibitem{Litim:2018} 
D.~F.~Litim and M.~J.~Trott,
``Asymptotic safety of scalar field theories,''
Phys.\ Rev.\ D {\bf 98} (2018) no.12, 125006.
doi:10.1103/PhysRevD.98.125006
[arXiv:1810.01678 [hep-th]].

\bibitem{Wetterich:1992yh} 
C.~Wetterich,
Phys.\ Lett.\ B {\bf 301}, 90 (1993).
doi:10.1016/0370-2693(93)90726-X
[arXiv:1710.05815 [hep-th]].

\bibitem{Reuter:1993kw} 
M.~Reuter and C.~Wetterich,
Nucl.\ Phys.\ B {\bf 417}, 181 (1994).
doi:10.1016/0550-3213(94)90543-6

\bibitem{Reuter:2019book}
M.~Reuter and F.~Saueressig
``Quantum Gravity and the Functional Renormalization Group: The Road towards Asymptotic Safety,''
Cambridge monographs on mathematical physics.
doi:10.1017/9781316227596.


\bibitem{Bonanno:2000ep}
A.~Bonanno, M.~Reuter,
``Renormalization group improved black hole spacetimes,''
Phys.\ Rev.\ D {\bf 62}, no. 10, 043008 (2000). 
doi:10.1103/PhysRevD.62.043008
[arXiv:hep-th/0002196].

\bibitem{Bonanno:2001xi}
A.~Bonanno, M.~Reuter,
``Cosmology of the Planck era from a renormalization group for quantum gravity,''
Phys.\ Rev.\ D {\bf 65}, 043508 (2002).
doi:10.1103/PhysRevD.65.043508
[arXiv:hep-th/0106133].

\bibitem{D.litim:2001} 
D. F. Litim,
``Optimized renormalization group flows,''
Phys.\ Rev.\ D {\bf 64}, 105007 (2001).
doi:10.1103/PhysRevD.64.105007

\bibitem{D.litim:2000plb} 
D. F. Litim,
``Optimisation of the exact renormalisation group,''
Phys.\ Lett.\ B 486 (2000) 92.
doi:10.1016/S0370-2693(00)00748-6

\bibitem{Reuter:2001ag} 
M.~Reuter and F.~Saueressig,
``Renormalization group flow of quantum gravity in the Einstein-Hilbert truncation,''
Phys.\ Rev.\ D {\bf 65}, 065016 (2002).
doi:10.1103/PhysRevD.65.065016

\bibitem{Beesham:1994ni}
A.~Beesham,
``Bianchi type I cosmological models with variable $G$ and $\Lambda$,''
Gen. Rel. Grav. {\bf 26}, 159 (1994). 
doi:10.1007/BF02105151.

\bibitem{kalligas:1995}
D.~Kalligas, P.~S.~Wesson and C.~W.~F.~Everitt,
``Bianchi type I cosmological models with variable $G$ and $\Lambda$: A comment,''
Gen. Rel. Grav. {\bf27}: 645 (1995).
doi:10.1007/BF02108066.

\bibitem{singh:2008as}
J.~P.~Singh, A.~Pradhan, A.~K.~Singh,
``Bianchi Type-I Cosmological Models with Variable G and $\Lambda$-Terms in General Relativity,''
Astrophys.SpaceSci.314:83-88 (2008). 
doi:10.1007/s10509-008-9742-6
[arXiv:0705.0459 [gr-qc]].

\bibitem{pradhan:2013as}
A.~Pradhan, R.~Jaiswal, R.~K.~Khare,
``Bianchi type-I cosmological models with time dependent $q$ and $\Lambda$-term in general relativity,''
Astrophys.Space Sci. 343 (2013) 489-497. 
doi:10.1007/s10509-012-1239-7.


\bibitem{kalligas:1992}
D.~Kalligas, P.~S.~Wesson and C.~W.~F.~Everitt,
``Flat FRW Models with Variable $G$ and $\Lambda$,''
Gen. Rel. Grav. {\bf24}: 351 (1992). 
doi:10.1007/BF00760411.












\end{thebibliography}
\end{document}